\documentclass[twocolumn,preprintnumbers,amsmath,amssymb]{revtex4}
\catcode`\@=11
\def\eqlt{\mathrel{\mathpalette\@vereq<}}  
\def\eqgt{\mathrel{\mathpalette\@vereq>}}  
\def\@vereq#1#2{\lower2.5pt\vbox{\baselineskip0pt \lineskip-.5pt
 \ialign{$\m@th#1\hfil##\hfil$\crcr#2\crcr{=}\crcr}}}

\newcommand{\simle}{\ \raise.3ex\hbox{$<$}\kern-0.8em\lower.7ex\hbox{$\sim$}\ }
\newcommand{\simge}{\ \raise.3ex\hbox{$>$}\kern-0.8em\lower.7ex\hbox{$\sim$}\ }

\newcommand{\mib}[1]{\mbox{\boldmath $#1$}}

\begin{document}
\title {Quantum Phase Transitions to Charge Order and Wigner Crystal Under Interplay of Lattice Commensurability and Long-Range Coulomb Interaction}  
\author {  Yohei Noda and Masatoshi  Imada }  
\address {Institute for Solid State Physics, University of Tokyo, Kashiwanoha,
Kashiwa, Chiba, 277-8581, Japan}  

\begin{abstract} 
Relationship among Wigner crystal, charge order and Mott insulator is studied
by the path-integral renormalization group method for two-dimensional lattices 
with long-range Coulomb interaction.  In contrast to Hartree-Fock 
results, the solid stability drastically increases with lattice 
commensurability. The transition to liquid occurs at the electron gas parameter $r_s \sim 2$ 
for the filling $n=1/2$ showing large reduction from $r_s \sim 35$ in the 
continuum limit.  Correct account of quantum fluctuations are crucial to understand charge-order 
stability generally observed only at simple fractional fillings and nature of quantum liquids 
away from them. 
\end{abstract}
\maketitle
Coulomb interaction between electrons drives various types of electron crystallization
ranging from Wigner lattice, charge order including stripes to 
Mott insulator.  
The Wigner transition in the continuum space was studied on two-dimensional (2D)
electron systems by quantum Monte Carlo calculations and the transition point 
was estimated to be
$r_s\sim 35$ at zero temperature~\cite{ImadaTakahashi,TanatarCeperley},
where $r_s=r_0/a_B$ with the Wigner Seitz radius $r_0=1/\sqrt{\pi n_e}$, 
the Bohr radius $a_B=4\pi\epsilon \hbar^2/m^*e^2$ and the electron density $n_e$. 
This 2D Wigner transition in quantum region was experimentally realized in GaAs
heterostructure with extremely high mobility and the transition at $r_s \sim 35$ was in
good agreement with the theoretical prediction~\cite{Shayegan}.

The most dramatic effect of atomic periodic potential in crystal not contained
in the electron gas is apparently
the band formation in electronic spectra due to Bloch theorem.  
In this letter, we stress another crucial
concept, lattice commensurability, generates different dramatic effects when the 
Coulomb interaction is present.   In correlated electron systems in solids as in 
transition metal 
compounds~\cite{ImadaFujimoriTokura,Tranquada} and organic systems~\cite{Kanoda},
charge ordering phenomena including stripe-type are common when the
electron filling $n$ satisfies simple fractional number such as 1/2 and 1/3.
These phenomena have attracted much interest in relation to 
the mechanisms of high-Tc superconductivity in the cuprates and 
colossal magnetoresistance in the manganites.  
However, in those systems, as we discuss later, the effective value of
$r_s$ is usually estimated to be rather small in the range $r_s<10$,
though we have some uncertainty in the evaluation of the effective mass $m^*$
and the dielectric constant $\epsilon$.  At such low values of $r_s$, we 
normally expect the quantum melting of electrons in terms of the above electron gas
picture.
When the electron filling deviates 
from a simple fractional number, 
such charge orders indeed usually melts to metals or the charge periodicity is pinned at 
the simple fillings with added carriers localized by disorder.  
It implies that a mechanism of 
stabilizing the solid only at simple commensurate filling works very efficiently.

Another known result is that the Mott insulator at integer fillings can be stabilized even at 
lower effective $r_s$. The Mott insulator can be viewed as the 
extreme limit of strong commensurability. A common feature of the charge
order and the Mott insulator is 
that they are charge imcompressible state resulted from the short-ranged part of
the Coulomb interaction.  
 
In this letter, we show that the mechanism for such common stabilization 
of the solid phase only at simple fractional fillings is not clear in the Hartree-Fock(HF) calculations,
whereas the stabilization and melting can indeed be understood only when the quantum
fluctuation effects are seriously considered.

To understand the basic mechanism, we employ a Hamiltonian given as 
\begin{eqnarray}
\hat{H} &=& \hat{H}_{0} + \hat{H}_{I} ,\nonumber\\ 
\hat{H}_{0} &=&
-\sum_{<i,j>,\sigma}\,t_{ij}(c_{i\sigma}^{\dagger}c_{j\sigma}+h.c.),
 \nonumber\\
\hat{H}_{I} &=& U\sum_{i}n_{i\uparrow}n_{i\downarrow} +
 \frac{1}{2}\sum_{i\ne j}V_{ij}n_{i}n_{j}, \label{Hamiltonian-spin} 
\end{eqnarray}
where the notations follow the standard one in the Hubbard type models.
We take the long-ranged Coulomb term as 
$V_{ij}=V/\vert r_i-r_j \vert$, where we ignore possible 
screening effects arising from electrons in other bands. The jelium model is employed by assuming a uniform 
positive charge as a background while the effect of ionic periodic potential is
represented by the lattice with single-band electrons located
near the Fermi level.   
The onsite interaction $U$ measured from $V$ may depend on the
detailed structure of atomic orbitals.
Here we take $U/t=4$ throughout this paper for simplicity.  This ratio
may be more or less the lower bound for the real situations. 
In any case our results in this letter for $n\le 1/2$ do not sensitively depend on this 
ratio. We study spinless fermion models in addition to
spin-1/2 electrons.   For spinless fermions, the $U$ term does not exist. This corresponds to
the fully polarized ferromagnetic case. 
We impose the periodic boundary condition in the numerical studies. The Coulomb interactions between images across the periodic  boundaries are taken by the Ewald summation.

Since our purpose in this letter is to understand a generic and universal feature of the 
interplay between quantum fluctuations and commensurability, 
we restrict our study to the square lattice with the anisotropy of lattice constants 
$a_x$ and $a_y$ as well as
 $t_x$ and $t_y$ taken as simple as possible to avoid a possible complexity which may intervene
the clarification of the essence.
The transfer is limited to the 
nearest-neighbor pairs in $x$ and $y$ directions denoted by $t_x$ and $t_y$ as in Fig.1.
We choose the anisotropy $d=a_y/a_x$ to make the charge ordering on a right triangular 
lattice possible.  By this choice, the charge
order stability is optimized and can be estimated on the same ground as the continuum limit. 
To see a systematic dependence on $r_s$, we control $t_x$ and $t_y$ by a 
single parameter of the effective mass $m^*$ as 
$t_z=\hbar^2/2m^*a_z^2$ with $z=x$ or $y$, which correctly reproduces the 
dispersion in the continuum limit. 

The above Hamiltonian may be rewritten as 
$\hat{H_R} = \hat{H}_{0} + \hat{H}_{I},
\hat{H}_{0} = -\frac{d}{\pi n r_s^2}\sum_{<i,j>_x,\sigma}
(c_{i\sigma}^{\dagger}c_{j\sigma}+h.c.)
-\frac{1}{\pi d n r_s^2}\sum_{<i,j>_y,\sigma}
(c_{i\sigma}^{\dagger}c_{j\sigma}+h.c.) 
+ \frac{2}{\pi n r_s^2}(d + \frac{1}{d})\hat{N}_e ,
\hat{H}_{I} = U\sum_{i}n_{i\uparrow}n_{i\downarrow} +
\sqrt{\frac{d}{\pi n}}\frac{1}{r_s}\sum_{i ,j}
\frac{n_{i}n_{j}}{|\mib{r}_j - \mib{r}_i|}$
in the energy unit of Rydbergs, where $\hat{N}_e$ is the total 
electron-number operator.
In the continuum limit, by
the limit of vanishing $a_x$ and $a_y$ and $n
\rightarrow 0$, with $r_s$ fixed, the above 
Hamiltonian is reduced to that of conventional electron gas.

%
%
%
%
%
\catcode`\@=11
%
%
\def\psfortextures{
\def\PSspeci@l##1##2{%
\special{illustration ##1\space scaled ##2}%
}}
\def\psfordvitops{
\def\PSspeci@l##1##2{%
\special{dvitops: import ##1\space \the\drawingwd \the\drawinght}%
}}
\def\psfordvips{
\def\PSspeci@l##1##2{%
\d@my=0.1bp \d@mx=\drawingwd \divide\d@mx by\d@my%
\includegraphics{##1\space}%
}}
\def\psforoztex{
\def\PSspeci@l##1##2{%
\special{##1 \space
      ##2 1000 div dup scale
      \putsp@ce{\number-\psllx} \putsp@ce{\number-\pslly} translate
}%
}}
\def\putsp@ce#1{#1 }
\def\psfordvitps{
\def\psdimt@n@sp##1{\d@mx=##1\relax\edef\psn@sp{\number\d@mx}}
\def\PSspeci@l##1##2{%
\special{dvitps: Include0 "psfig.psr"}
\psdimt@n@sp{\drawingwd}
\special{dvitps: Literal "\psn@sp\space"}
\psdimt@n@sp{\drawinght}
\special{dvitps: Literal "\psn@sp\space"}
\psdimt@n@sp{\psllx bp}
\special{dvitps: Literal "\psn@sp\space"}
\psdimt@n@sp{\pslly bp}
\special{dvitps: Literal "\psn@sp\space"}
\psdimt@n@sp{\psurx bp}
\special{dvitps: Literal "\psn@sp\space"}
\psdimt@n@sp{\psury bp}
\special{dvitps: Literal "\psn@sp\space startTexFig\space"}
\special{dvitps: Include1 "##1"}
\special{dvitps: Literal "endTexFig\space"}
}}
\def\psonlyboxes{
\def\PSspeci@l##1##2{%
\at(0cm;0cm){\boxit{\vbox to\drawinght
  {\vss
  \hbox to\drawingwd{\at(0cm;0cm){\hbox{(##1)}}\hss}
  }}}
}%
}
\def\psloc@lerr#1{%
\let\savedPSspeci@l=\PSspeci@l%
\def\PSspeci@l##1##2{%
\at(0cm;0cm){\boxit{\vbox to\drawinght
  {\vss
  \hbox to\drawingwd{\at(0cm;0cm){\hbox{(##1) #1}}\hss}
  }}}
\let\PSspeci@l=\savedPSspeci@l
}%
}
%
%
\newread\psiz@
\newdimen\drawinght\newdimen\drawingwd
\newdimen\psxoffset\newdimen\psyoffset
\newbox\drawingBox
\newif\ifNotB@undingBox
\newhelp\PShelp{Proceed: you'll have a 5cm square blank box instead of
your graphics (Jean Orloff).}
\def\@mpty{}
\def\s@tsize#1 #2 #3 #4\@ndsize{
  \def\psllx{#1}\def\pslly{#2}%
  \def\psurx{#3}\def\psury{#4}
  \ifx\psurx\@mpty\NotB@undingBoxtrue
  \else
    \drawinght=#4bp\advance\drawinght by-#2bp
    \drawingwd=#3bp\advance\drawingwd by-#1bp
  \fi
  }
\def\sc@nline#1:#2\@ndline{\edef\p@rameter{#1}\edef\v@lue{#2}}
\def\g@bblefirstblank#1#2:{\ifx#1 \else#1\fi#2}
\def\psm@keother#1{\catcode`#112\relax}
\def\execute#1{#1}
{\catcode`\%=12
\xdef\B@undingBox{
}  		
\def\ReadPSize#1{
 \edef\PSfilename{#1}
 \openin\psiz@=#1\relax
 \ifeof\psiz@ \errhelp=\PShelp
   \errmessage{I haven't found your postscript file (\PSfilename)}
   \psloc@lerr{was not found}
   \s@tsize 0 0 142 142\@ndsize
   \closein\psiz@
 \else
   \loop
     \execute{\begingroup
       \let\do\psm@keother
       \dospecials
       \catcode`\ =10
       \catcode`\^^M=9
       \global\read\psiz@ to\n@xtline
       \endgroup}
     \ifeof\psiz@
       \errhelp=\PShelp
       \errmessage{(\PSfilename) is not an Encapsulated PostScript File:
           I could not find any \B@undingBox: line.}
       \edef\v@lue{0 0 142 142:}
       \psloc@lerr{is not an EPSFile}
       \NotB@undingBoxfalse
     \else
       \expandafter\sc@nline\n@xtline:\@ndline
       \ifx\p@rameter\B@undingBox\NotB@undingBoxfalse
         \edef\int@rmediateresult{%
           \expandafter\g@bblefirstblank\v@lue\space\space\space}
         \expandafter\s@tsize\int@rmediateresult\@ndsize
       \else\NotB@undingBoxtrue
       \fi
     \fi
   \ifNotB@undingBox\repeat
   \closein\psiz@
 \fi
\message{#1}
}
%
%
\newcount\xscale \newcount\yscale \newdimen\pscm\pscm=1cm
\newdimen\d@mx \newdimen\d@my
\let\ps@nnotation=\relax
\def\psboxto(#1;#2)#3{\vbox{
   \ReadPSize{#3}
   \divide\drawingwd by 1000
   \divide\drawinght by 1000
   \d@mx=#1
   \ifdim\d@mx=0pt\xscale=1000
         \else \xscale=\d@mx \divide \xscale by \drawingwd\fi
   \d@my=#2
   \ifdim\d@my=0pt\yscale=1000
         \else \yscale=\d@my \divide \yscale by \drawinght\fi
   \ifnum\yscale=1000
         \else\ifnum\xscale=1000\xscale=\yscale
                    \else\ifnum\yscale<\xscale\xscale=\yscale\fi
              \fi
   \fi
   \divide \psxoffset by 1000\multiply\psxoffset by \xscale
   \divide \psyoffset by 1000\multiply\psyoffset by \xscale
   \global\divide\pscm by 1000
   \global\multiply\pscm by\xscale
   \multiply\drawingwd by\xscale \multiply\drawinght by\xscale
   \ifdim\d@mx=0pt\d@mx=\drawingwd\fi
   \ifdim\d@my=0pt\d@my=\drawinght\fi
   \message{scaled \the\xscale}
 \hbox to\d@mx{\hss\vbox to\d@my{\vss
   \global\setbox\drawingBox=\hbox to 0pt{\kern\psxoffset\vbox to 0pt{
      \kern-\psyoffset
      \PSspeci@l{\PSfilename}{\the\xscale}
      \vss}\hss\ps@nnotation}
   \global\ht\drawingBox=\the\drawinght
   \global\wd\drawingBox=\the\drawingwd
   \baselineskip=0pt
   \copy\drawingBox
 \vss}\hss}
  \global\psxoffset=0pt
  \global\psyoffset=0pt
  \global\pscm=1cm
  \global\drawingwd=\drawingwd
  \global\drawinght=\drawinght
}}
%
%
\def\psboxscaled#1#2{\vbox{
  \ReadPSize{#2}
  \xscale=#1
  \message{scaled \the\xscale}
  \divide\drawingwd by 1000\multiply\drawingwd by\xscale
  \divide\drawinght by 1000\multiply\drawinght by\xscale
  \divide \psxoffset by 1000\multiply\psxoffset by \xscale
  \divide \psyoffset by 1000\multiply\psyoffset by \xscale
  \global\divide\pscm by 1000
  \global\multiply\pscm by\xscale
  \global\setbox\drawingBox=\hbox to 0pt{\kern\psxoffset\vbox to 0pt{
     \kern-\psyoffset
     \PSspeci@l{\PSfilename}{\the\xscale}
     \vss}\hss\ps@nnotation}
  \global\ht\drawingBox=\the\drawinght
  \global\wd\drawingBox=\the\drawingwd
  \baselineskip=0pt
  \copy\drawingBox
  \global\psxoffset=0pt
  \global\psyoffset=0pt
  \global\pscm=1cm
  \global\drawingwd=\drawingwd
  \global\drawinght=\drawinght
}}
%
\def\psbox#1{\psboxscaled{1000}{#1}}
%
%
\def\centinsert#1{\midinsert\line{\hss#1\hss}\endinsert}
\def\psannotate#1#2{\def\ps@nnotation{#2\global\let\ps@nnotation=\relax}#1}
\def\pscaption#1#2{\vbox{
   \setbox\drawingBox=#1
   \copy\drawingBox
   \vskip\baselineskip
   \vbox{\hsize=\wd\drawingBox\setbox0=\hbox{#2}
     \ifdim\wd0>\hsize
       \noindent\unhbox0\tolerance=5000
    \else\centerline{\box0}
    \fi
}}}
\def\psfig#1#2#3{\pscaption{\psannotate{#1}{#2}}{#3}}
\def\psfigurebox#1#2#3{\pscaption{\psannotate{\psbox{#1}}{#2}}{#3}}
%
\def\at(#1;#2)#3{\setbox0=\hbox{#3}\ht0=0pt\dp0=0pt
  \rlap{\kern#1\vbox to0pt{\kern-#2\box0\vss}}}
%
\newdimen\gridht \newdimen\gridwd
\def\gridfill(#1;#2){
  \setbox0=\hbox to 1\pscm
  {\vrule height1\pscm width.4pt\leaders\hrule\hfill}
  \gridht=#1
  \divide\gridht by \ht0
  \multiply\gridht by \ht0
  \gridwd=#2
  \divide\gridwd by \wd0
  \multiply\gridwd by \wd0
  \advance \gridwd by \wd0
  \vbox to \gridht{\leaders\hbox to\gridwd{\leaders\box0\hfill}\vfill}}
%
\def\fillinggrid{\at(0cm;0cm){\vbox{
  \gridfill(\ht\drawingBox;\wd\drawingBox)}}}
%
%
\def\textleftof#1:{
  \setbox1=#1
  \setbox0=\vbox\bgroup
    \advance\hsize by -\wd1 \advance\hsize by -2em}
\def\textrightof#1:{
  \setbox0=#1
  \setbox1=\vbox\bgroup
    \advance\hsize by -\wd0 \advance\hsize by -2em}
\def\endtext{
  \egroup
  \hbox to \hsize{\valign{\vfil##\vfil\cr%
\box0\cr%
\noalign{\hss}\box1\cr}}}
%
\def\frameit#1#2#3{\hbox{\vrule width#1\vbox{
  \hrule height#1\vskip#2\hbox{\hskip#2\vbox{#3}\hskip#2}%
        \vskip#2\hrule height#1}\vrule width#1}}
\def\boxit#1{\frameit{0.4pt}{0pt}{#1}}
\catcode`\@=12 
%
 \psfordvips   

\begin{figure}
$$ \psboxscaled{400}{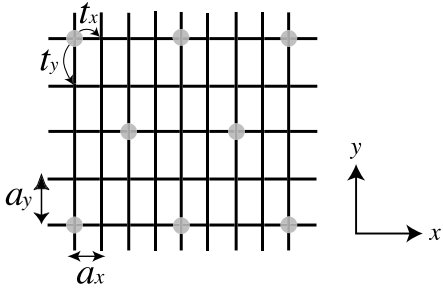} $$
\caption{Lattice structure with anisotropic transfers
$t_x$ and $t_y$}
\label{Fig1}
\end{figure}

\begin{figure}
$$ \psboxscaled{500}{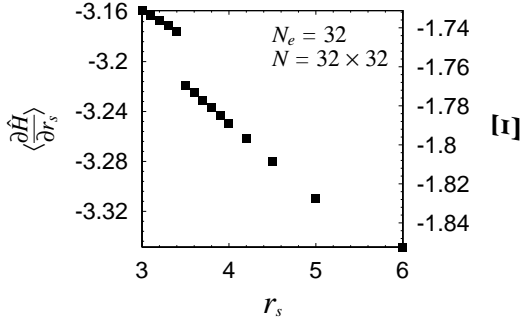} $$
\caption{Hartree-Fock result for $dH/dr_s$ vs. $r_s$ for 32 spinless electrons on 32 by 32 lattices.}
\label{JumpHF}
\end{figure}

 We first show results of the standard HF calculation for spinless fermions.
For the fillings
$n<1/2$, the first-order transitions are clearly visible from the jump
of $\langle d\hat{H}/dr_s\rangle$ e.g. in Fig.~\ref{JumpHF} at $r_{sc}=3.45\pm0.05$ for $n=1/32$ on $32 \times 32$
lattice. Note that the jump of $\langle d\hat{H}/dr_s \rangle$
indicates the level crossing in the ground state as a function of the control parameter $r_s$
and gives an evidence for the first-order transition. 
The right ordinate is the scale for the second Hamiltonian $\hat{H}_R$ defined by 
$\Xi= \partial (r_s^2\hat{H}_{R})/\partial r_s =
(d/\pi n) \partial \hat{H}/\partial r_s. 
$
From Fig.~\ref{TwoBodyCFHF},
the electron crystallization is indeed identified in the two-body correlation
function defined by 
$C(\mib{r_i})= \frac{1}{N}\,\sum_{\mib{r_j}} \big\langle
n_{\mib{r_j+r_i}}n_{\mib{r_j}} \big\rangle.
$

\begin{figure}
$$ \psboxscaled{550}{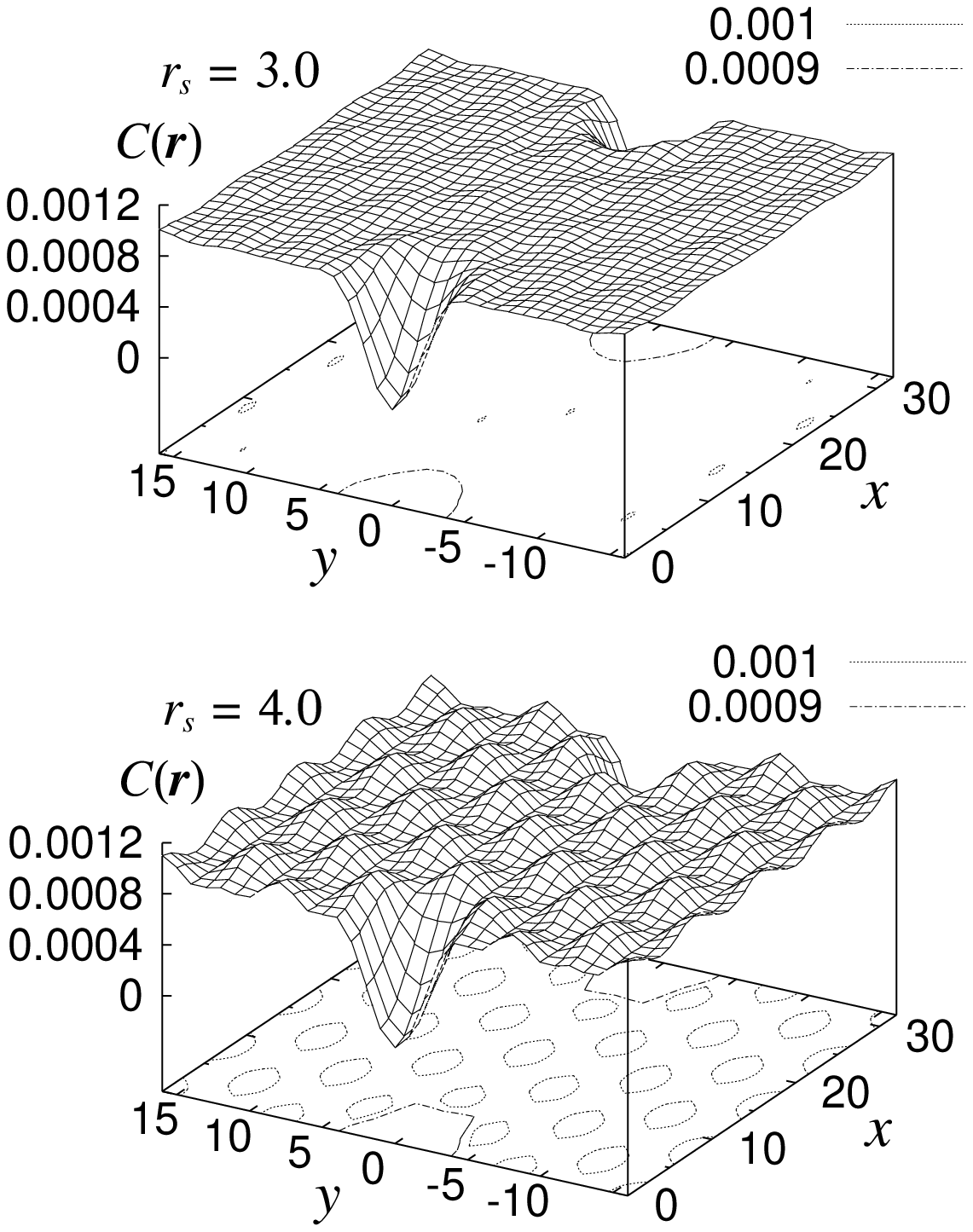} $$
\caption{Hartree-Fock result for two-body correlation function for 32 spinless electrons on 
32 by 32 lattice.}
\label{TwoBodyCFHF}
\end{figure}


For fixed electron number $N_e$, the transition point is extrapolated to the continuum limit
$n \rightarrow 0$. These results are further extrapolated to the thermodynamic limit $N_e 
\rightarrow \infty$ as in Fig.~\ref{ExtrapHF}.  
This indicates that the system size dependence is 
small and the result at $N_e=8$ is already a good estimate of the thermodynamic limit. 
At $n=1/2$, the transition is of continuous type with the absence of a jump in
$\langle dH/dr_s \rangle $ while the peak value of the
Fourier transform of two-body charge correlation diverges as a Bragg peak with increasing system size
for $r_s\ge r_{sc}$.  Then the transition point is estimated as $r_{sc}=0.9\pm 0.1$ after 
finite-size scaling.

\begin{figure}
$$ \psboxscaled{500}{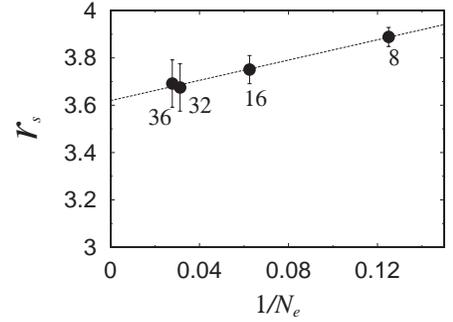} $$
\caption{The Hartree-Fock results for liquid-solid transition points of spinless electrons 
as a function of inverse of the particle number $N_e$ in the 
limit $n \rightarrow 0$.}
\label{ExtrapHF}
\end{figure}

To understand effects of spin degrees of freedom, spin-1/2 electrons are also analyzed at
$n=1/4$ and $1/2$.  The transitions appear to be of first order at $r_{sc}=2.05 \pm 0.05$ 
and continuous  at $r_{sc}=1.35 \pm 0.15$, respectively. The orders of transitions are the same as 
the spinless case and the difference of $r_{sc}$ between these two  cases
shrinks with decreasing filling. We discuss this tendency later.  We conclude that the unrestricted HF approximation predicts 
the crystallization of 2D electrons ranging from $r_{sc}\sim 1$ 
at $n=1/2$ to $r_{sc}\sim 3.69$ in the continuum limit. We note that the filling dependence 
is rather small. 
   
To understand how quantum and many-body fluctuations modifies the 
HF results, 
we have calculated the ground state of this system by applying recently developed
PIRG method~\cite{KashimaImadaPIRG}.  
This algorithm allows us to start from and improve the above unrestricted HF solutions and 
reach the correct ground state by taking account quantum fluctuations in a controlled fashion.
By following the path integral formalism, the convergence to the ground state is obtained after repeated operations of $\exp[-\tau \hat{H}]$
to Slater determinants with small $\tau$ and by the 
break-up of the kinetic and Coulomb energy terms. The long-ranged Coulomb term 
is rewritten by using the Stratonovich-Hubbard transformation which allows this term to operate
to the Slater determinants, though it increases the number of Slater determinants in the 
linear combination of the wave function~\cite{NodaImada}.
With increasing the dimension of truncated Hilbert space in a nonorthogonal 
basis optimized by this path integral operation, an extrapolation method using the variance
$\Delta E=(\langle \hat{H}^2 \rangle - \langle \hat{H} \rangle^2)/\langle \hat{H} \rangle^2$, 
is employed to 
reach the true ground state. For the 
extrapolation, we took the number of nonorthogonal Slater
determinants up to 256.  
We
 show results 
when the extrapolation linearly converges well to those of the full Hilbert space.  

We took lattice sizes up to 144 sites ($12 \times 12$ lattice)
with electrons up to 72 at quarter filling $n=1/2$ for both
spinless and spin-1/2 electron systems (Then for 144 sites of spin-1/2 system, we took 36 up and down spin 
electrons each). For fillings 1/8 and 1/18, we show only on the spinless system for 64 sites 
and 144 sites, respectively with 8 electrons.  
We restricted to the spinless case for 
these low fillings since the difference of $r_{sc}$
from the system with spins is estimated to be small. This is inferred from the comparisons of the PIRG with the HF results at 
$n=1/2$ together with the HF results at lower filing with 
spins. In fact, the spin effects appear mainly 
through the exchange process. The exchange is scaled by $t^2$ where $t$ is 
the effective transfer between neighboring electrons. At lowering fillings, it forms lower-energy hierarchy 
as compared with that of the charge order, $V$. Therefore, though spin is important in 
determining the magnetic order, 
this small energy scale hardly changes the solid-liquid transition point for $n \le 1/2$.
From comparisons for different sizes, the system-size dependence of the 
transition point appears to be small at the low fillings because the 
transition becomes first order. Therefore, though our 
systems at the low fillings are rather small, the results are expected to be close to the thermodynamic results.

We first show spinless cases. 
Fig.~\ref{JumpPIRG} shows a typical example showing the first-order transition in the PIRG calculation
at $n=1/18$ for 8 fermions on 12 by 12 lattice.
The behavior of jump is similar to the HF results with a clear change of the two-body correlation function
as in Fig.~\ref{TwoBodyCFPIRG}. However, this first-order crystallization takes place
at substantially higher $r_s$ than the Hartre-Fock result.  The transition at $n=1/8$ is also
first order while that is of continuous type again at $n=1/2$.  The transition point is estimated to be
$r_{sc}= 1.75 \pm 0.25, 13.5 \pm 0.5$ and $24.5 \pm 0.5$ for $n=1/2, 1/8$ and 1/18, respectively.    
We have also studied spin-1/2 electrons at $n=1/2$ and obtained a continuous transition at $r_{sc}
=2.0\pm 1.0$, which is comparable to the spinless case.  

\begin{figure}
$$ \psboxscaled{580}{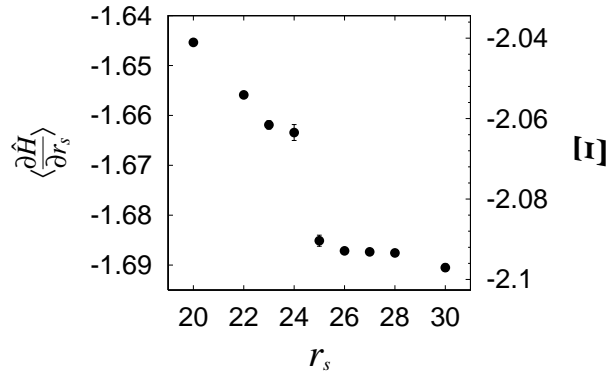} $$
\caption{PIRG result for $dH/dr_s$ vs. $r_s$ for 8 spinless electrons on 12 by 12 lattices.}
\label{JumpPIRG}
\end{figure}

\begin{figure}
$$ \psboxscaled{550}{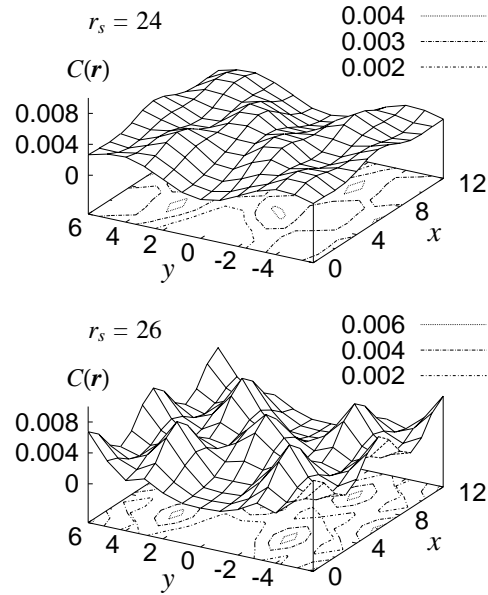} $$
\caption{PIRG result for two-body correlation function for 8 spinless electrons on 
12 by 12 lattice.}
\label{TwoBodyCFPIRG}
\end{figure}
\begin{figure}
$$ \psboxscaled{600}{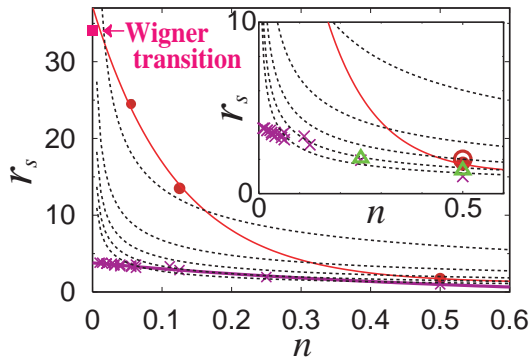} $$
\caption{Solid-liquid phase boundary in plane of the filling $n$ and $r_s$.
The square indicates the Wigner transition point~\cite{ImadaTakahashi}. 
The PIRG results (filled and open circles for spinless and spin-1/2 fermions, respectively) are compared with the Hartree-Fock results (crosses for spinless and triangles for spin-1/2 electrons). The solid curves are guides for the eye. The dotted curves show contour lines for the ratio of $\epsilon/m^*$ to the bare value being at 1,2,3,4 and 5 from up to down.  The inset enlarges small $r_s$ region.}
\label{PhaseDiagram}
\end{figure}

In Fig.~\ref{PhaseDiagram}, the phase boundary of the solid and liquid phases, $r_{sc}$, are shown and the HF and PIRG results are compared for $n=1/l$ with integer $l$.  
At simple fractional filling such as $n=1/2$, $r_{sc}$ by the HF approximation
is relatively good. However, it deviates from the PIRG results rapidly with increasing
$l$.  The ratio of these two estimates increases from $\sim 1.5-2$ at $n=1/2$ to $\sim 10$ in the 
continuum limit. In terms of the electron density, this difference means from 3-4 to 100, 
since the density is scaled by $r_s^2$.  
From the PIRG results, we can understand why charge orders are
commonly observed in organic and transition metal compounds at simple 
fractional fillings as 1/2 and 1/3 while it is pinned or melts away from such
fillings because $r_{sc}$ increases dramatically with increasing denominator of 
the irreducible fraction while the parameters 
of many compounds may lie in this range of variation. For example, when we assume a 
layered perovskite structure with the lattice constant $a=4 \AA$ and the effective ratio
$\epsilon/m^*$ being the twice of the bare value, $n=1/2, 1/3$ and 2/3 are located in the solid region  
while $n=1/4, 3/4$ and fractions with higher denominators are located in the liquid region
as we see in Fig.~\ref{PhaseDiagram}. We note that the filling control 
at fixed lattice constants follows a contour line of $\epsilon/m^*$ in Fig.~\ref{PhaseDiagram}. 
As we also see in Fig.~\ref{PhaseDiagram}, the HF approximation fails in accounting this general
experimental trend. 
The general trend can be understood only 
by taking account of quantum fluctuations sufficiently.

Even metallic states near the 
charge order may inherently contain such 
charge-order and stripe proximities e.g. formation of quantum and dynamical 
defect structure, in the region where the static solid could be stabilized 
on the mean-field level.
Competitions with
hierarchical structure formation by various levels of fractional numbers of fillings become intrinsic  
properties leading to self-organized structure.
This may be particularly conspicuous near the simple commensurate order such as the Mott 
insulator, because the mass renormalization increases effective $r_s$
with more continuous character of the transition. 
Reanalyses of mechanisms of high-Tc superconductivity in the cuprates and the colossal magnetoresistance in the manganites from the present viewpoint would be desired in future, since
they appear in the proximity of insulator stabilized by the commensurability.

In summary, we have studied the liquid-solid transition of two-dimensional electrons with long-range
Coulomb interaction.  We have shown
how the Mott insulator at $n=1$, charge orders at $0<n<1$ and the Wigner lattice at $n \rightarrow 0$
are connected. The charge order transition does not sensitively depend on the spin degrees of 
freedom for $n <1/2$.  Although the HF transition points
are rather insensitive to the electron filling and lattice commensurability, the PIRG 
result with quantum fluctuations taken in a correct way shows rapid increase of $r_s$ at the transition 
 with decreasing filling.  
The present PIRG results explain why charge orders or stripes are commonly observed only 
at simple fractional fillings such as 1/2 and 1/3 but usually are not away from them. 
  We stress that such interplay between quantum fluctuations and the commensurability may play crucial role also in metals (quantum liquids).
It would be interesting if the interplay can be experimentally studied by microfabrication of periodic potential to high-mobility 2D electron systems.

The work is supported by JSPS under grant JSPS-RFTF97P01103. A part of computation was done at the supercomputer center at ISSP, University of Tokyo. 
     
\end{document}